\documentclass[prb,nofootinbib,longbibliography,twocolumn,floatfix]{revtex4-2}
\usepackage{amsmath}
\usepackage{amssymb}
\usepackage{graphicx}
\usepackage{dcolumn}
\usepackage{bm}
\usepackage{microtype}
\usepackage[colorlinks=true,linkcolor=blue,anchorcolor=blue, citecolor=cyan,urlcolor=cyan]{hyperref}
\usepackage[mathlines]{lineno}
\usepackage{ulem}
\usepackage{xcolor}
\ifdefined\cleanversion
  
  \colorlet{red}{black}
\else
  
\fi
\makeatletter
\newenvironment{appendixfigure}{%
  \par\addvspace{\intextsep}\noindent
  \begin{minipage}{\columnwidth}\def\@captype{figure}\centering
}{%
  \end{minipage}\par\addvspace{\intextsep}
}
\makeatother
\begin{document}
\title{Steering and Local Tunneling Signatures of Majorana Corner Modes in Altermagnetic Josephson Junctions}
\author{Yu-Xuan Li$^{1,2}$}
\author{Tao Zhou$^{1,2}$}
\email{tzhou@scnu.edu.cn}
\affiliation{$^1$Guangdong Basic Research Center of Excellence for Structure and Fundamental Interactions of Matter, Guangdong Provincial Key Laboratory of Quantum Engineering and Quantum Materials, School of Physics, South China Normal University, Guangzhou 510006, China}
\affiliation{$^2$Guangdong-Hong Kong Joint Laboratory of Quantum Matter, Frontier Research Institute for Physics, South China Normal University, Guangzhou 510006, China}

\begin{abstract}
	Dynamical manipulation of Majorana zero modes and their reliable distinction from topologically trivial states remain paramount challenges in topological quantum computation. Here, we propose a phase-biased altermagnetic Josephson junction as a versatile platform for generating and controlling Majorana corner configurations. Our platform utilizes the macroscopic superconducting phase difference and the N\'eel-vector orientation as independent control knobs to reshape the boundary mass. This allows the spatial configuration of Majorana corner modes (MCMs) to be selected at different corners of a fixed device geometry. This spatial reconfiguration yields a control-correlated conductance response. Across these configurations, the quantized zero-bias peak appears at the contacted corner in the ideal tunneling model when an MCM is present and is suppressed when it is absent. This correlation connects phase and magnetic control of corner configurations with local tunneling readout.
\end{abstract}

\maketitle
\section{Introduction}
Controllable manipulation and reliable identification of Majorana zero modes (MZMs) remain the central challenges in topological superconductivity (TSC) and the essential prerequisites for non-Abelian quantum operations~\cite{qi_topological_2011,alicea_new_2012,sato_topological_2017,fu_superconducting_2008,alicea_majorana_2010,lutchyn_majorana_2010,sato_topological_2009,sau_generic_2010}. While conventional one-dimensional TSCs suffer from stray-field perturbations induced by necessary magnetic elements~\cite{alicea_majorana_2010,lutchyn_majorana_2010,oreg_helical_2010}, higher-order topological superconductors offer a compelling alternative. By binding MZMs to the corners of a two-dimensional finite sample, they provide natural spatial separation, which facilitates site-selective detection and topological braiding~\cite{langbehn_reflection-symmetric_2017,hsu_majorana_2018,wang_high-temperature_2018,khalaf_symmetry_2018,yan_majorana_2018,zhu_tunable_2018,zhang_helical_2019,zhang_topological_2020,pahomi_braiding_2020,zhou_phase_2020,pan_detecting_2022,wu_boundary-obstructed_2020,lee_odd-parity_2021,pan_majorana_2024}. Furthermore, the recent emergence of altermagnets~\cite{Libor_Beyond_2022,mazin_editorial_2022,smejkal_emerging_2022,zhang_finite_momentum_2024,xiao_spin_2024,bai_altermagnetism_2024,liu_altermagnetism_2025,fukaya_superconducting_2025,Jungwirth2026}---featuring strong non-relativistic spin splitting without net macroscopic magnetization---provides a promising foundation for TSC~\cite{zhu_topological_2023,ghorashi_altermagnetic_2024,Zhu_Dislocation_2024,Pal_jose_2025,sun2025,fz2b-5sp7,PhysRevB.111.064502,65q6-5wxl,Sharma_Doublepeak_2026,fu_Floquet_2026,yang_topological_altermagnetic_2026,zhuAltermagneticProximityEffect2026,mcardle_2026,li_spin_polarized_2026}. Consequently, higher-order TSC has been theoretically predicted in altermagnetic heterostructures, giving rise to Majorana corner modes (MCMs) whose localization can be manipulated by the N\'eel-vector orientation or strain~\cite{li_majorana_2023,li_realizing_2024,Qiao_MCM_2025,hadjipaschalisMajoranas2025}. Toward prospective non-Abelian operations, dynamic braiding protocols in these platforms have also begun to be explored~\cite{hodge_altermagnet-superconductor_2025}.

\begin{figure}[t]
	\includegraphics[width=0.95\linewidth]{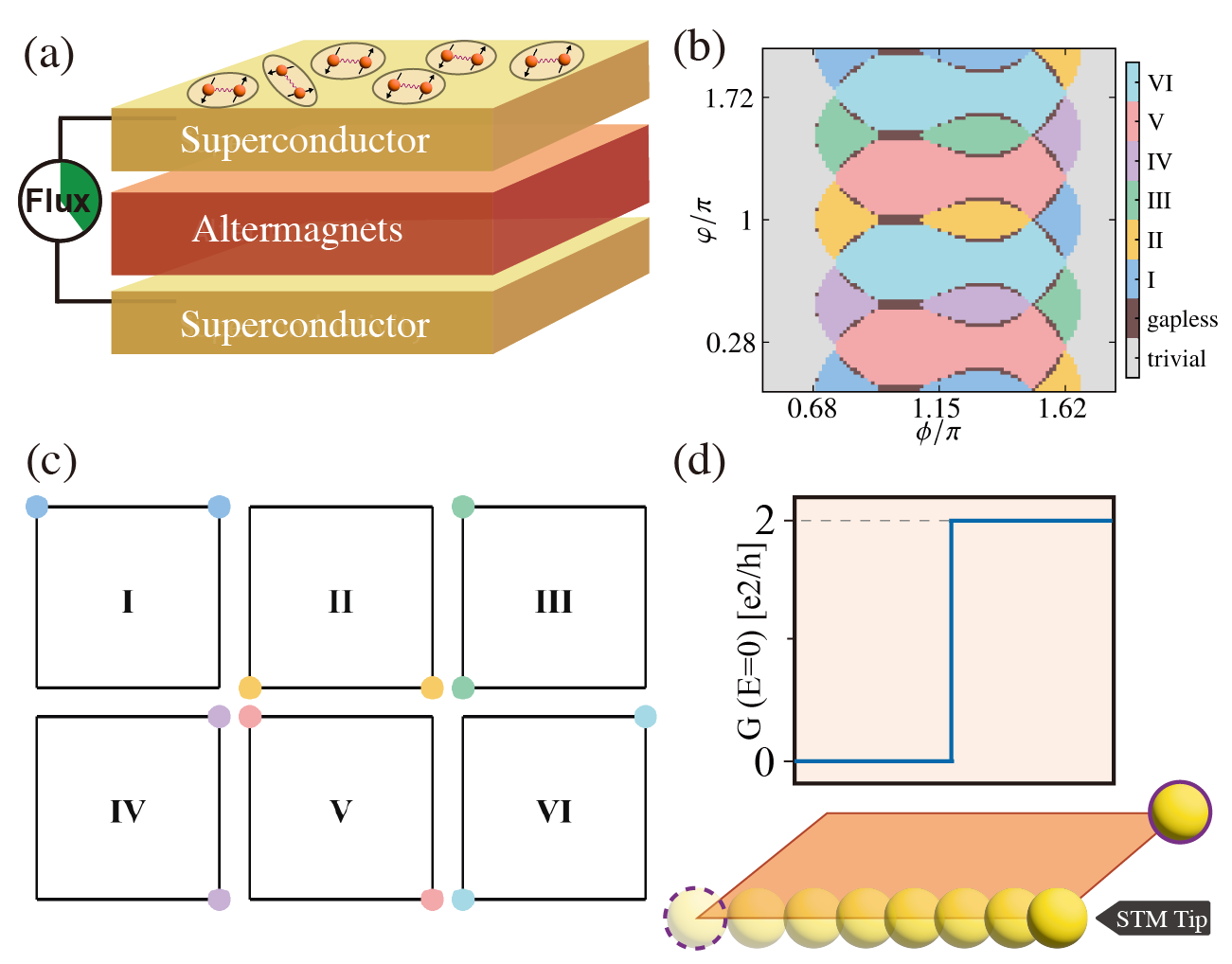}
	\caption{
		(a) Schematic of the phase-biased altermagnetic Josephson junction, where the superconducting phase difference $\phi$ is controlled by the magnetic flux.
		(b) Phase diagram in the $(\phi,\varphi)$ plane, with $\varphi$ the N\'eel-vector azimuthal angle. Regions I--VI denote distinct MCM corner-localization regimes; brown and gray regions indicate gapless and trivial regimes, respectively.
		(c) Corner-localization patterns of the MCMs for the six regions labeled in (b). Colored dots mark the corners occupied by MCMs.
		(d) Local tunneling readout of the zero-bias conductance near a contacted corner.
	}
	\label{fig:illustrate}
\end{figure}

Despite these theoretical milestones, developing an altermagnetic scheme that combines spatial control with local detection faces two challenges. First, true non-Abelian braiding~\cite{ivanov_non-abelian_2001,nayak_non-abelian_2008} requires the deterministic, non-destructive spatial movement of MZMs. Existing proposals often rely on tuning material parameters or constructing domain-wall configurations, which can complicate control of the mode locations within a fixed device~\cite{alicea_wire_2011,aasen_milestones_2016}. Second, the standard experimental signature of MZMs, an isolated zero-bias conductance peak (ZBCP), can also arise from topologically trivial Andreev bound states (ABS)~\cite{liuAndreevBoundStates2017,prada_andreev_2020}. Such states can mimic a solitary Majorana signal, rendering static ZBCP measurements ambiguous and motivating spatially resolved, control-correlated measurements.

In this work, we propose a controllable approach to spatial control and local readout of MCMs by designing a phase-biased altermagnetic Josephson junction [Fig.~\ref{fig:illustrate}(a)]. We demonstrate that this architecture serves as a controllable platform for reconfiguring Majorana corner modes. By incorporating an altermagnet as the weak link between conventional superconducting electrodes, we exploit the interplay between the macroscopic superconducting phase difference $\phi$ and the N\'eel-vector azimuthal angle $\varphi$ as two parameters for controlling the corner-mode distribution.

We reveal that traversing the $(\phi, \varphi)$ parameter space systematically reshapes the boundary mass, subdividing the higher-order topological phase space into distinct corner-localization regimes [Regions I--VI in Fig.~\ref{fig:illustrate}(b)]. Rather than merely switching the corner modes on or off, varying these parameters selects different MCM corner pairs within a fixed device geometry [Fig.~\ref{fig:illustrate}(c)]. This spatial reconfiguration produces a corresponding change in the local conductance. As illustrated in Fig.~\ref{fig:illustrate}(d), the local tunneling response tracks the relocation of the MCMs. In the ideal tunneling model, the $2e^2/h$ zero-bias peak appears at a contacted corner hosting an MCM and is suppressed when the MCM moves away. Measurements at different corners and control settings thus connect the conductance pattern with the predicted MCM configuration, providing more information than an isolated static ZBCP. Our results establish this junction as a platform for Majorana corner control and local readout.

\section{Model}
For the junction geometry shown in Fig.~\ref{fig:illustrate}(a), we use an effective-bilayer model for two conducting channels associated with the upper and lower interfaces. We include interfacial Rashba spin--orbit coupling to capture structural inversion asymmetry, with opposite signs at the two interfaces. The insulating altermagnet is represented by a proximity-induced exchange field rather than explicit low-energy states. The electronic states are described in the layer-spin basis $c_{\boldsymbol{k}}
=
\left(
c_{1\uparrow,\boldsymbol{k}},
c_{1\downarrow,\boldsymbol{k}},
c_{-1\uparrow,\boldsymbol{k}},
c_{-1\downarrow,\boldsymbol{k}}
\right)^T ,$
where $\ell=\pm1$ labels the two layers, and the Pauli matrices $\ell_i$ and $s_i$ act on layer and spin spaces, respectively.
In this basis, the normal-state Hamiltonian is
\begin{equation}
	\begin{aligned}
		h(\boldsymbol{k})
		=&\,
		\xi_{\boldsymbol{k}}\ell_0s_0
		+
		\alpha
		\left(
		k_y s_x-k_xs_y
		\right)\ell_z
		+
		\Gamma \ell_xs_0
		\\
		&+
		J_0 f_\beta(\boldsymbol{k})
		\ell_0
		\boldsymbol{n}\cdot \boldsymbol{s},
	\end{aligned}
	\label{eq:h_normal_compact}
\end{equation}
where $\xi_{\boldsymbol{k}}=\hbar^2 k^2/(2m)-\mu$, and $\alpha$ is the Rashba coupling. In the following analytical discussion and numerical calculations, we focus on the representative point $\mu=0$.
The parameter $\Gamma$ denotes  interlayer tunneling. 
The last term represents the altermagnetic spin splitting, with amplitude $J_0$, N\'eel vector direction $\boldsymbol n$, and momentum dependence
\begin{equation}
	f_\beta(\boldsymbol{k})
	=
	\cos(\beta)
	\left(k_x^2-k_y^2\right)
	+
	\sin(\beta)k_xk_y ,
	\label{eq:f_beta_continuum}
\end{equation}
where $\beta$ specifies the orientation of this $d$-wave spin splitting in momentum space. 
The limits $\beta=0$ and $\beta=\pi/2$ correspond to the $B_{1g}$ $(d_{x^2-y^2})$ and $B_{2g}$ $(d_{xy})$ forms, respectively~\cite{Jungwirth2026}.
The direction of the N\'eel vector is parameterized as
\begin{equation}
	\boldsymbol{n}
	=
	(
	\sin\theta\cos\varphi,\,
	\sin\theta\sin\varphi,\,
	\cos\theta
	),
	\label{eq:n_vector}
\end{equation}
where $\theta$ and $\varphi$ denote its polar and azimuthal angles, respectively.

Superconductivity is introduced through layer-dependent $s$-wave pairing potentials induced by two parent superconductors. 
We take the pairing amplitude in the $\ell=1$ layer to be real and positive, $\Delta_1=|\Delta_1|$, while the pairing amplitude in the $\ell=-1$ layer is written as $\Delta_{-1}=|\Delta_{-1}|e^{i\phi}$. 
The phase difference $\phi$ can be tuned by embedding the junction in a superconducting loop, where fluxoid quantization relates the gauge-invariant phase bias to the enclosed magnetic flux~\cite{golubov_current_phase_2004,kayyalha_skewed_2020}. The pairing matrix in layer-spin space is
\begin{equation}
	\hat{\Delta}
	=
	i s_y
	\left(
	\Delta_1 P_+
	+
	\Delta_{-1}P_-
	\right),
	\qquad
	P_\pm=\frac{\ell_0\pm\ell_z}{2}.
	\label{eq:pairing_matrix}
\end{equation}
We treat the electrode-induced singlet pairing in the conducting layers as an effective parameter. Its proximity origin and the accompanying triplet correlations are analyzed in the Supplemental Material~\cite{supp}.

For boundary calculations, we employ a lattice-regularized form that preserves its long-wavelength structure and permits open boundary conditions.
We consider a ribbon geometry, periodic along one in-plane direction and finite along the other, to calculate the edge states, and a fully finite geometry to resolve the real-space localization and relocation of the MCMs~\cite{supp}.

\section{Topological phases at $\phi=\pi$}
We first identify the parent TSC phase in the absence of altermagnets.
At the $\pi$-junction configuration, the induced pairing potentials in the two layers have opposite signs,
$\Delta_{1} = -\Delta_{-1}\equiv \Delta_0$.
The resulting BdG Hamiltonian preserves time-reversal symmetry $\mathcal{T}=is_y\mathcal{K}$
and particle-hole symmetry $\mathcal{P}=\gamma_x\mathcal{K}$, placing the system in class DIII~\cite{schnyder_classification_2008,supp}.
Crucially, the $\pi$-phase bias ensures that the Hamiltonian commutes with the mirror operator
$M_z = \gamma_0 s_z \ell_x$.
This allows a block-diagonalization into two mirror sectors, as detailed in Appendix~\ref{app:mirror} and
Supplemental Material~\cite{supp}.

For $\mu=0$, the bulk gap closes at $\boldsymbol{k}=0$ when $|\Gamma| = |\Delta_0|$,
signaling the transition between trivial ($|\Gamma| < |\Delta_0|$) and
topological ($|\Gamma| > |\Delta_0|$) regimes.
In the topological regime, each mirror sector realizes a chiral TSC and supports a single chiral
Majorana boundary mode.
The two mirror sectors are related by time reversal, which reverses the chirality.
Consequently, the full $\phi = \pi$ system hosts a pair of counterpropagating Majorana edge modes,
forming helical Majorana edge states.

An explicit boundary-state calculation gives the corresponding low-energy edge theory
(see Appendix~\ref{app:edge}).
Projecting the continuum BdG Hamiltonian onto the two time-reversal-related Majorana edge modes yields
\begin{equation}
	H_{\rm edge}^{(0)}
	=
	v k_\parallel \chi_z ,
	\label{eq:helical_edge}
\end{equation}
where $k_{\parallel}$ denotes the momentum along the boundary and $\chi_i$ act in the subspace spanned
by the helical Majorana edge modes.
Equation~(\ref{eq:helical_edge}) describes a pair of counterpropagating Majorana edge states,
confirming that the $\phi = \pi$ junction realizes a helical TSC phase.


\section{Altermagnetism-induced MCMs}
 Having established the helical TSC phase at $\phi=\pi$, we next examine the response of the helical Majorana edge modes to altermagnetic spin splitting at $\phi=\pi$.
For the $B_{1g}$ $(d_{x^2-y^2})$ spin splitting $(\beta=0)$, out-of-plane and in-plane N\'eel vector orientations have qualitatively different effects on the helical edge states.

\begin{figure}
	\includegraphics[width=\linewidth]{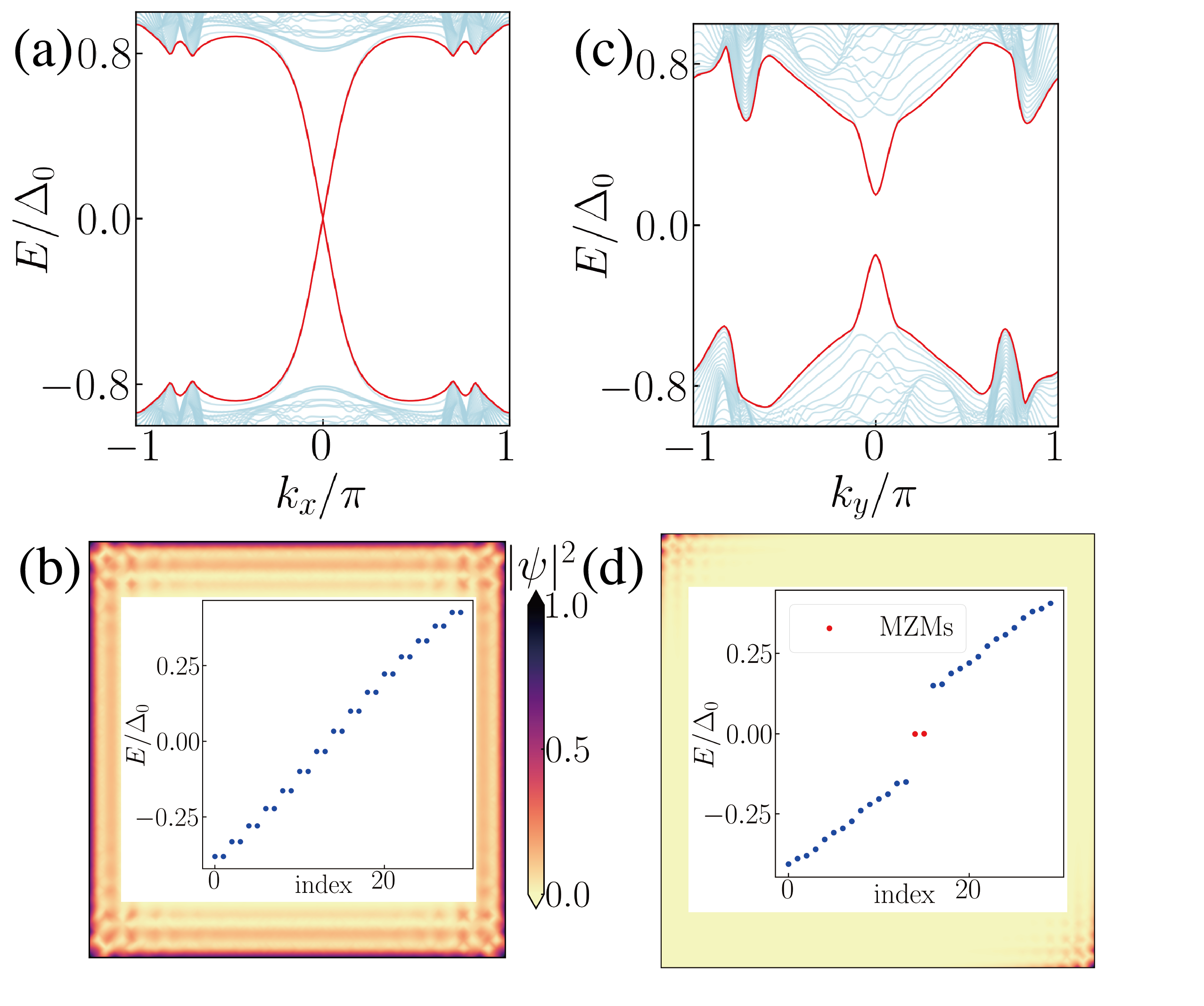}
	\caption{
		Boundary spectra and real-space distributions for different N\'eel vector orientations.
		(a),(b) For $\boldsymbol n\parallel\hat z$, the cylinder spectrum hosts gapless helical Majorana edge modes, whose lowest-energy states extend along the boundaries in a fully finite geometry.
		(c),(d) For $\boldsymbol n\parallel[11]$, the altermagnetic spin splitting gaps the boundary modes and produces zero-energy modes localized at the corners.
		The insets in (b) and (d) show the corresponding finite-size spectra.
	}
	\label{fig:mcms}
\end{figure}

For an out-of-plane N\'eel vector orientation, $\boldsymbol n\parallel\hat z$, the $B_{1g}$ $(d_{x^2-y^2})$ spin splitting leaves the helical edge states gapless, as shown in Fig.~\ref{fig:mcms}(a).
The finite-size calculation in Fig.~\ref{fig:mcms}(b) is consistent with this boundary structure.
The low-energy wave-function weight is distributed along the sample boundary, and the finite-size spectrum exhibits discrete in-gap levels characteristic of a closed one-dimensional Majorana boundary channel.
For a boundary of length $L$, the low-energy levels are approximately quantized as $E_n\simeq \pm v(2n+1)\pi/L$, corresponding to an antiperiodic boundary condition for the Majorana edge mode.
The absence of a boundary gap follows from the mirror-sector structure of the helical edge states~\cite{supp}.
For $\boldsymbol n\parallel\hat z$, the spin-splitting term preserves the mirror symmetry $M_z$ and therefore cannot hybridize the counterpropagating Majorana modes from opposite mirror sectors.

A qualitatively different response occurs for an in-plane N\'eel vector.
For the in-plane orientation $\boldsymbol n\parallel[11]$ considered here, the spin-splitting term couples the counterpropagating Majorana edge states and opens a finite boundary gap, as shown in Fig.~\ref{fig:mcms}(c).
In a fully finite geometry, two zero-energy states remain isolated inside the gap [inset of Fig.~\ref{fig:mcms}(d)], with probability densities localized at the two corners connected by the $[1\bar{1}]$ diagonal.
The origin of these corner states is captured by the edge theory~\cite{supp}.
After projection onto the helical edge subspace, the spin splitting generates a boundary mass whose sign depends on the edge orientation.
For the configuration in Fig.~\ref{fig:mcms}(d), the projected boundary mass reverses sign at the two corners connected by the $[1\bar{1}]$ diagonal~\cite{supp}, so that these corners act as mass domain walls binding MZMs~\cite{jackiwSolitonsFermionNumber1976}. The zero-energy spectra and corner-localized wave functions, together with the boundary mass sign reversal, establish these bound states as MCMs.

\begin{figure}
	\includegraphics[scale=0.46]{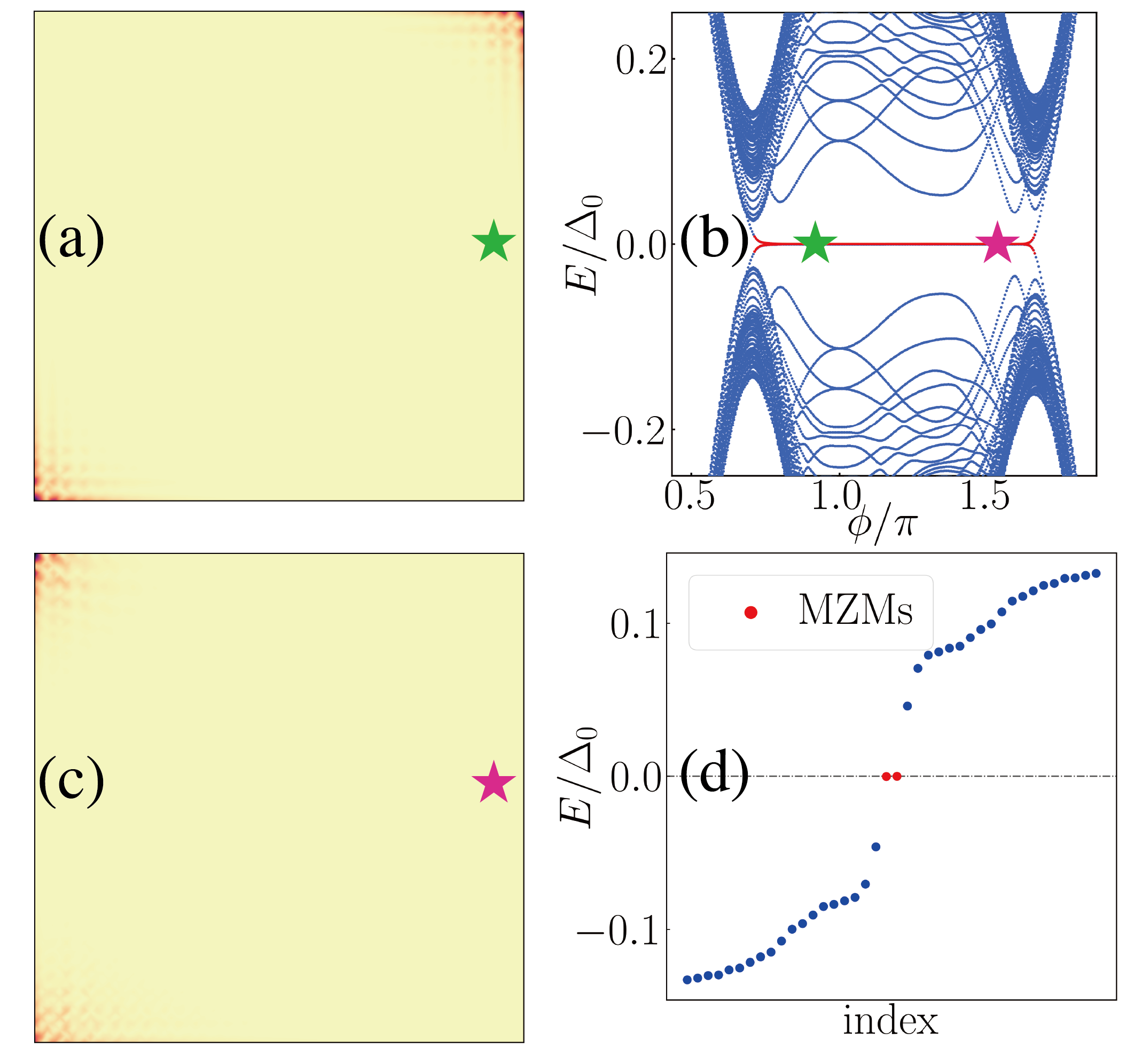}
	\caption{
		Control and phase robustness of MCMs.
		(a) Probability density of the MCMs at $\phi=\pi$ and $\boldsymbol n\parallel[1\bar{1}]$.
		(b) Finite-size spectrum versus $\phi$, demonstrating the persistence of the zero-energy modes over a finite phase interval.
		(c),(d) Probability density and corresponding finite-size spectrum away from the $\pi$-junction limit at $\varphi=0.35\pi$, showing corner-localized MCMs isolated within the quasiparticle gap.
	}
	\label{fig:phi}
\end{figure}

\section{Control of MCMs}
The corner configuration of the MCMs can be controlled by the N\'eel-vector orientation and the superconducting phase difference.
We first illustrate the N\'eel vector  control for the $B_{1g}$ $(d_{x^2-y^2})$ spin splitting $(\beta=0)$ at $\phi=\pi$.
For a fixed phase difference, the in-plane N\'eel vector   orientation determines the sign pattern of the  boundary mass~\cite{supp}.
When $\boldsymbol n\parallel[11]$, the mass domain walls localize the MCMs at the two corners connected by the $[1\bar{1}]$ diagonal.
Changing the N\'eel vector to $\boldsymbol n\parallel[1\bar{1}]$ reverses the boundary mass configuration and places the MCMs at the two corners along the $[11]$ diagonal, as shown in Fig.~\ref{fig:phi}(a).
Thus, the N\'eel vector  orientation controls which diagonal pair of corners hosts the MCMs.

The superconducting phase difference provides a complementary tuning parameter for the MCM localization. To examine the stability away from the $\pi$-junction condition, we fix an in-plane N\'eel vector orientation and calculate the finite-size spectrum as a function of $\phi$, as shown in Fig.~\ref{fig:phi}(b).
The zero-energy modes persist over a finite phase interval around $\phi=\pi$ and remain separated from the higher-energy quasiparticle spectrum.
Thus, the MCMs do not require exact phase tuning to $\phi=\pi$, but survive under moderate deviations of the superconducting phase difference.

The superconducting phase difference can further reshape the spatial localization of the MCMs.
Figures~\ref{fig:phi}(c,d) show a representative configuration with N\'eel vector azimuthal angle $\varphi=0.35\pi$ and $\phi\neq\pi$.
The finite-size spectrum contains two zero-energy states isolated inside the bulk gap [Fig.~\ref{fig:phi}(d)], and the corresponding probability density is concentrated near the two corners on the left boundary [Fig.~\ref{fig:phi}(c)].
Together with the diagonal configuration in Fig.~\ref{fig:phi}(a), these results illustrate the different corner arrangements accessible in the $(\phi,\varphi)$ parameter space within the same finite geometry.
Taken together, the N\'eel vector orientation and the Josephson phase difference provide tunable parameters for controlling the corner localization of the MCMs, which motivates the local tunneling probe discussed below.


\begin{figure}
	\includegraphics[width=\linewidth]{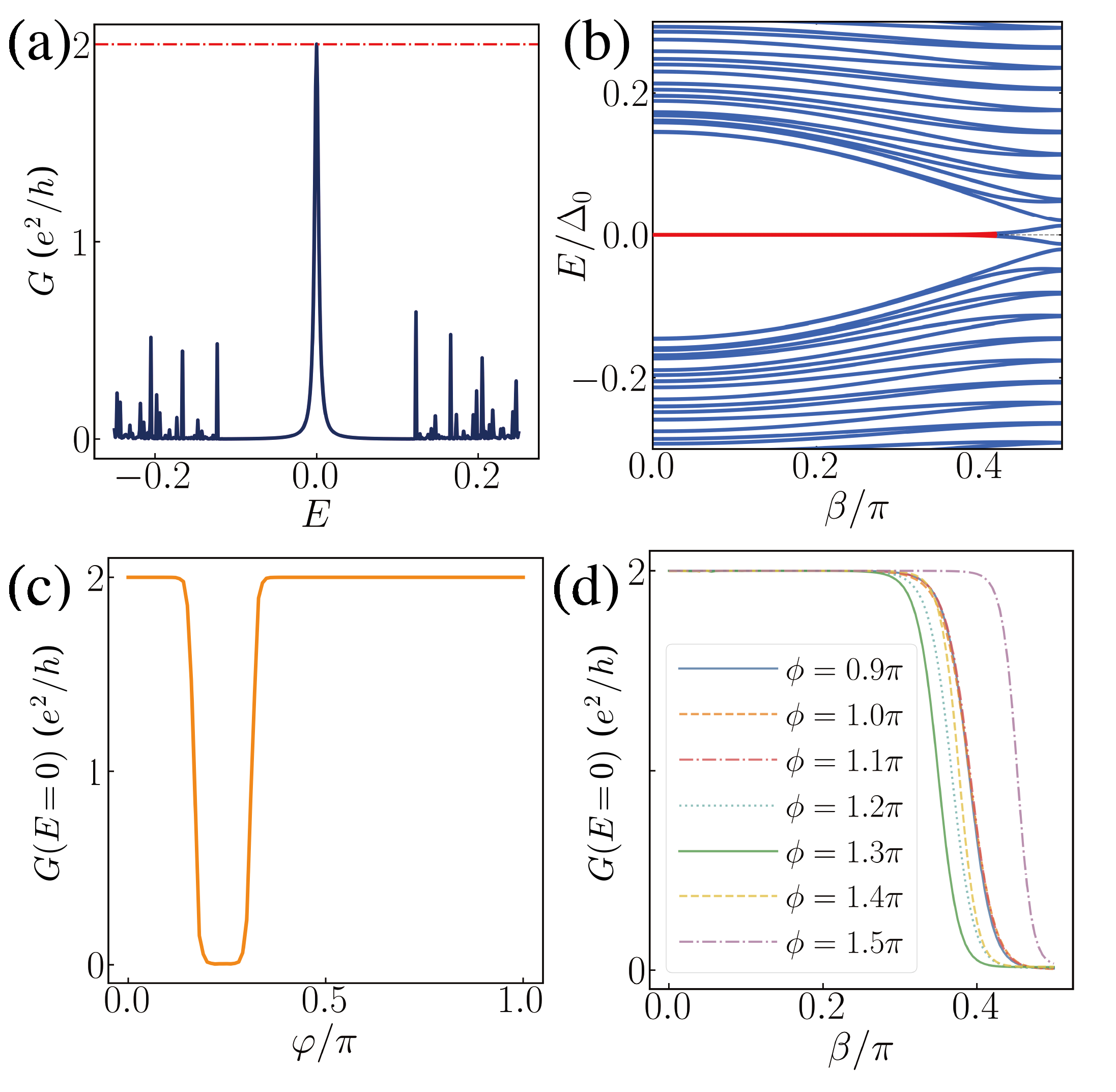}
	\caption{
	(a) Energy-resolved differential conductance $G(E)$ measured near a contacted corner.
		(b) Finite-size spectrum as a function of the form-factor parameter $\beta$.
		(c) Zero-bias conductance $G(0)$ as a function of the N\'eel vector azimuthal angle $\varphi$.
		(d) Zero-bias conductance $G(0)$ versus $\beta$ for representative superconducting phase differences $\phi$.
	}
	\label{fig:transport}
\end{figure}

\section{Local signature of MCMs}

The tunability of the MCM positions motivates a local spectroscopic probe of their relocation.
To this end, we attach a normal-metal lead near a corner of the finite system and calculate the differential conductance within the Blonder--Tinkham--Klapwijk scattering formalism~\cite{blonder_transition_1982,anantramCurrentFluctuationsMesoscopic1996}. At zero temperature, the conductance is given by
\begin{equation}
	G(E)
	=
	\frac{e^2}{h}
	\left[
	N_m(E)-R_{ee}(E)+R_{he}(E)
	\right],
	\label{eq:conductance}
\end{equation}
where $R_{ee}={\rm Tr}(r^{ee\dagger}r^{ee})$ and $R_{he}={\rm Tr}(r^{he\dagger}r^{he})$ denote the normal- and Andreev-reflection probabilities, respectively, and $N_m(E)$ is the number of propagating electron modes in the normal lead.
In the ideal zero-temperature limit, an MCM coupled to the probe produces resonant Andreev reflection with a quantized conductance $G(0)=2e^2/h$~\cite{law_majorana_2009}.

Figure~\ref{fig:transport}(a) shows the energy-resolved conductance for a configuration in which the contacted corner hosts an MCM.
The Majorana-mediated resonant Andreev reflection produces a sharp zero-bias peak reaching $2e^2/h$, whereas the finite-bias conductance remains nonuniversal.
This quantized zero-bias response provides the reference signal for the calculated MCM locally coupled to the tunneling probe.

We further analyze the dependence of the MCMs on the form-factor parameter $\beta$ and the superconducting phase difference $\phi$, which respectively characterize the symmetry of the altermagnets and the deviation from the $\pi$-junction condition.
The parameter $\beta$ interpolates between the $B_{1g}$ $(d_{x^2-y^2})$ and $B_{2g}$ $(d_{xy})$ form factors of the spin splitting.
As shown in Fig.~\ref{fig:transport}(b), the zero-energy modes persist over a finite interval of $\beta$, indicating that the MCMs are not tied to an ideal $B_{1g}$ spin-splitting form but remain stable in the presence of a $B_{2g}$ admixture.
The phase difference $\phi$ probes the stability of the MCMs away from the $\pi$-junction condition, where the helical parent TSC and the effective edge theory are most naturally established. The zero-bias conductance in Fig.~\ref{fig:transport}(d) shows that the MCM response persists over finite ranges of both $\phi$ and $\beta$.
For a fixed tunneling contact, $G(0)$ approaches $2e^2/h$ when an MCM is localized at the contacted corner, whereas the response is strongly suppressed once the MCMs are relocated away from that corner.
Thus, both the zero-energy modes and their local tunneling signatures are robust against the studied deviations from the pure $B_{1g}$ spin splitting and from the $\pi$-junction condition.

The dependence on the N\'eel vector azimuthal angle provides a complementary test of the corner-resolved tunneling response.
As shown in Fig.~\ref{fig:transport}(c), the zero-bias conductance varies strongly with $\varphi$.
The conductance remains close to $2e^2/h$ when the MCM configuration places a zero mode at the contacted corner, and is suppressed when the MCMs occupy other corners.
Together with the phase-dependent response in Fig.~\ref{fig:transport}(d), this behavior demonstrates that the local tunneling conductance follows the MCM redistribution driven by both the N\'eel-vector orientation and the superconducting phase difference. The $\varphi$- and $\phi$-dependent conductance responses therefore establish a direct correspondence between the controlled MCM relocation and the spatial evolution of the local zero-bias signal.
For the tested helical triplet channel, the corner-mode configurations persist over $0\leq\eta/\Delta\leq0.4$. Changing the N\'eel-vector azimuth from $\varphi=-\pi/4$ to $+\pi/4$ continues to switch the favored corner diagonal, although the corner contrast decreases as $\eta$ increases~\cite{supp}.

\section{Experimental considerations}
The proposed platform combines established approaches for controlling
the superconducting phase difference $\phi$ and the N\'eel vector. The phase bias can be tuned by embedding the junction in a flux-threaded superconducting loop, as demonstrated in proximitized Josephson structures~\cite{fornieri_evidence_2019,ren_phasecontrolled_2019,kayyalha_skewed_2020}. The junction current--phase response must be included when calibrating the actual phase difference~\cite{supp}.

In this phase-biased geometry, $\phi$ is an externally constrained thermodynamic variable rather than an unconstrained parameter. Its equilibrium value is determined by minimizing the total free energy of the junction and the external circuit. Although the isolated junction favors $\phi=0$ for the parameters used in the auxiliary calculation~\cite{supp}, this minimum does not determine the operating phase of the externally biased device. The fixed-$\phi$ calculations therefore describe a junction coupled to phase-biased superconducting reservoirs.

The N\'eel vector orientation can be manipulated by uniaxial strain~\cite{liebman_strain_2026}, while supercurrent-driven N\'eel torques offer a complementary theoretical possibility~\cite{vakili_supercurrent_2026}. Applying strain control to the proposed stack requires sufficient magnetoelastic coupling and compatibility with superconducting operation.

A possible implementation uses an insulating altermagnet such as rutile NiF$_2$ as a spacer between the conducting Rashba layers~\cite{hariki_separating_2025,shi_exchange_2004}. NiF$_2$ has a well-defined in-plane N\'eel easy axis, and epitaxial (110) thin films have been grown, with reported film N\'eel temperatures near 81 K~\cite{hariki_separating_2025,shi_exchange_2004}. Its insulating character allows it to supply effective exchange without adding low-energy itinerant bands to the spacer. The superconducting electrodes induce pairing in the conducting layers. The phase bias may be controlled with an asymmetric SQUID, and the N\'eel vector may be reoriented by strain transferred from a piezoelectric substrate or actuator. The spacer and interfaces must permit exchange transfer and interlayer tunneling while retaining the required relative Rashba-coupling sign and sufficiently uniform magnetic order. Together, these material and control conditions provide a route to selecting distinct stationary MCM configurations and their associated local conductance responses within a fixed device geometry.

\section{Summary}
In summary, we have established phase-biased altermagnetic Josephson junctions as a controllable platform for topological quantum physics. Moving beyond static measurements, our approach exploits the interplay between altermagnetism and superconductivity to select Majorana corner configurations. The resulting spatially correlated conductance switching provides a local readout of the calculated configurations, rather than an exclusive Majorana signature. Importantly, we have verified that these zero modes and their local tunneling signatures exhibit robustness over the studied finite intervals of the phase difference and the altermagnetic form-factor, without requiring exact $\pi$ bias or a pure $B_{1g}$ form factor.

\begin{acknowledgments}
	This work was supported by the 
	Guangdong Provincial Quantum Science Strategic Initiative under Grant No. GDZX2601002 and the National Natural Science
	Foundation of China under Grant No. 12604242.
\end{acknowledgments}

\appendix
\raggedbottom

\section{Mirror-symmetry block diagonalization}
\label{app:mirror}
The unperturbed Rashba bilayer Josephson junction, biased at phase difference $\phi=\pi$ and under the continuum limit (equal intra-layer pairing $\Delta_0$), is governed by the $8\times 8$ Bogoliubov--de Gennes (BdG) Hamiltonian
\begin{equation}
H_\pi(\boldsymbol{k}) = \varepsilon_{\boldsymbol{k}}\gamma_z + \alpha(k_y s_x - k_x s_y\gamma_z)\ell_z + \Gamma\ell_x\gamma_z + \Delta_0\ell_z\gamma_y s_y.
\label{eq:H_pi_full}
\end{equation}
Here, $\gamma_i, s_i, \ell_i$ are Pauli matrices operating in Nambu, spin, and layer subspaces, respectively. The normal-state dispersion is $\varepsilon_{\boldsymbol{k}} = k^2/(2m) - \mu$. We set $\hbar=1$ in this appendix and use the common Nambu gauge $G=\gamma_z$, so the pairing block is $-\hat\Delta$ relative to the convention in Sec.~II. This does not change the relative superconducting phase.

At $\phi=\pi$, Eq.~(\ref{eq:H_pi_full}) commutes with the mirror operator $M_z=\gamma_0s_z\ell_x$ ($M_z^2=1$). The unitary transformation $U$ specified in the Supplemental Material includes the reordering into mirror eigenspaces and the basis change within each sector. It gives $U^\dagger H_\pi(\boldsymbol{k})U=\operatorname{diag}[H_+(\boldsymbol{k}),H_-(\boldsymbol{k})]$, where
\begin{equation}
H_\pm(\boldsymbol{k}) = -\varepsilon_{\boldsymbol{k}}\lambda_z s_0 - \alpha k_x \lambda_0 s_x + \alpha k_y \lambda_z s_y \mp \Gamma\lambda_z s_z - \Delta_0\lambda_y s_y,
\end{equation}
with $\lambda_i$ acting on the newly defined particle-hole pseudo-subspace. Time reversal exchanges the sectors and reverses momentum, $V_T H_+^*(\boldsymbol{k})V_T^\dagger=H_-(-\boldsymbol{k})$, where $V_T=-\lambda_0s_y$ in this basis. A topological phase transition occurs when the bulk gap closes at $\boldsymbol{k}=0$, explicitly given by the condition $|\Gamma| = |\Delta_0|$ (assuming $\mu=0$). For $|\Gamma| > |\Delta_0|$, each sector enters a topological superconducting phase supporting a single chiral Majorana edge mode.

\section{Helical edge states and subspace projection}
\label{app:edge}
To obtain the effective edge Hamiltonian, we project each mirror sector onto its low-energy Dirac subspace near the bulk transition, as derived in the Supplemental Material. For a boundary at $y=0$, the sample occupies $y>0$, and the vacuum is represented by a trivial region with the opposite Dirac mass.

The resulting mass domain wall supports a normalizable edge state in each sector, with its spinor fixed by the decay condition. Combining the two sectors gives counterpropagating Majorana edge modes related by time reversal. Projection of the tangential momentum term onto these modes yields the effective low-energy Hamiltonian
\begin{equation}
H^{(0)}_{\text{edge}} = v k_\parallel \chi_z,
\end{equation}
where $v$ is the effective Fermi velocity, $k_\parallel$ is the momentum along the boundary tangent, and $\chi_i$ are Pauli matrices operating in the basis of these two helical edge modes. The projection operator $\mathcal{P}_\vartheta$ maps any bulk operator onto this low-energy helical subspace for a boundary oriented at angle $\vartheta$.

\section{Altermagnetic boundary mass and MCM steering}
\begin{appendixfigure}
\centering
 \includegraphics[width=\linewidth]{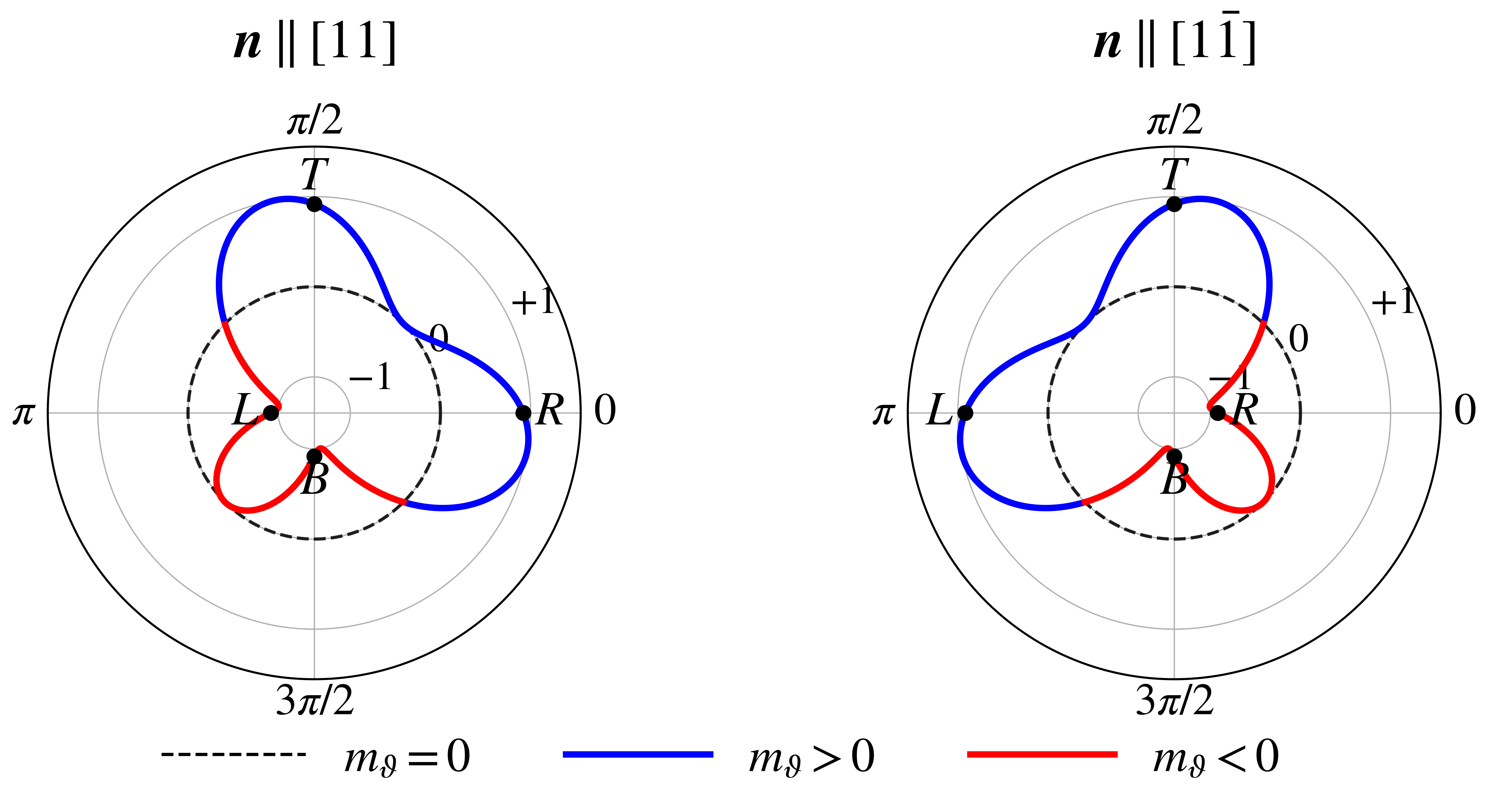} 
\caption{Angular dependence of the altermagnetic boundary mass $m_\vartheta$ on the outward-normal angle $\vartheta$, shown in polar coordinates for two N\'eel-vector orientations. Blue and red denote positive and negative boundary masses, respectively.}
\label{fig:mass_polar}
\end{appendixfigure}
We now introduce the altermagnetic bulk perturbation $H_{\text{AM}} = J_0 f_\beta(\boldsymbol{k}) \boldsymbol{n}\cdot\boldsymbol{s}$. For a specific boundary characterized by an outward normal $\hat{\boldsymbol{\nu}} = (\cos\vartheta, \sin\vartheta)$ and tangent $\hat{\boldsymbol{t}} = (-\sin\vartheta, \cos\vartheta)$, the spin splitting at the Dirac point is dominated by the transverse momentum, so that $f_\beta(\boldsymbol{k}) \rightarrow k_\perp^2 [ \cos\beta\cos(2\vartheta) + \frac{1}{2}\sin\beta\sin(2\vartheta) ]$.

Applying the subspace projection $\mathcal{P}_\vartheta$, we find that only the in-plane spin component parallel to the boundary tangent, $\boldsymbol{s}_\parallel \cdot \hat{\boldsymbol{t}}$, yields a non-vanishing off-diagonal coupling, with $\mathcal{P}_\vartheta (\boldsymbol{s}_\parallel \cdot \hat{\boldsymbol{t}}) \mathcal{P}_\vartheta \propto \chi_y$. For a N\'eel vector lying in the plane at azimuthal angle $\varphi$, its projection along the tangent is exactly $\sin(\varphi-\vartheta)$. 

\begin{appendixfigure}
	\centering
	\includegraphics[width=0.95\linewidth]{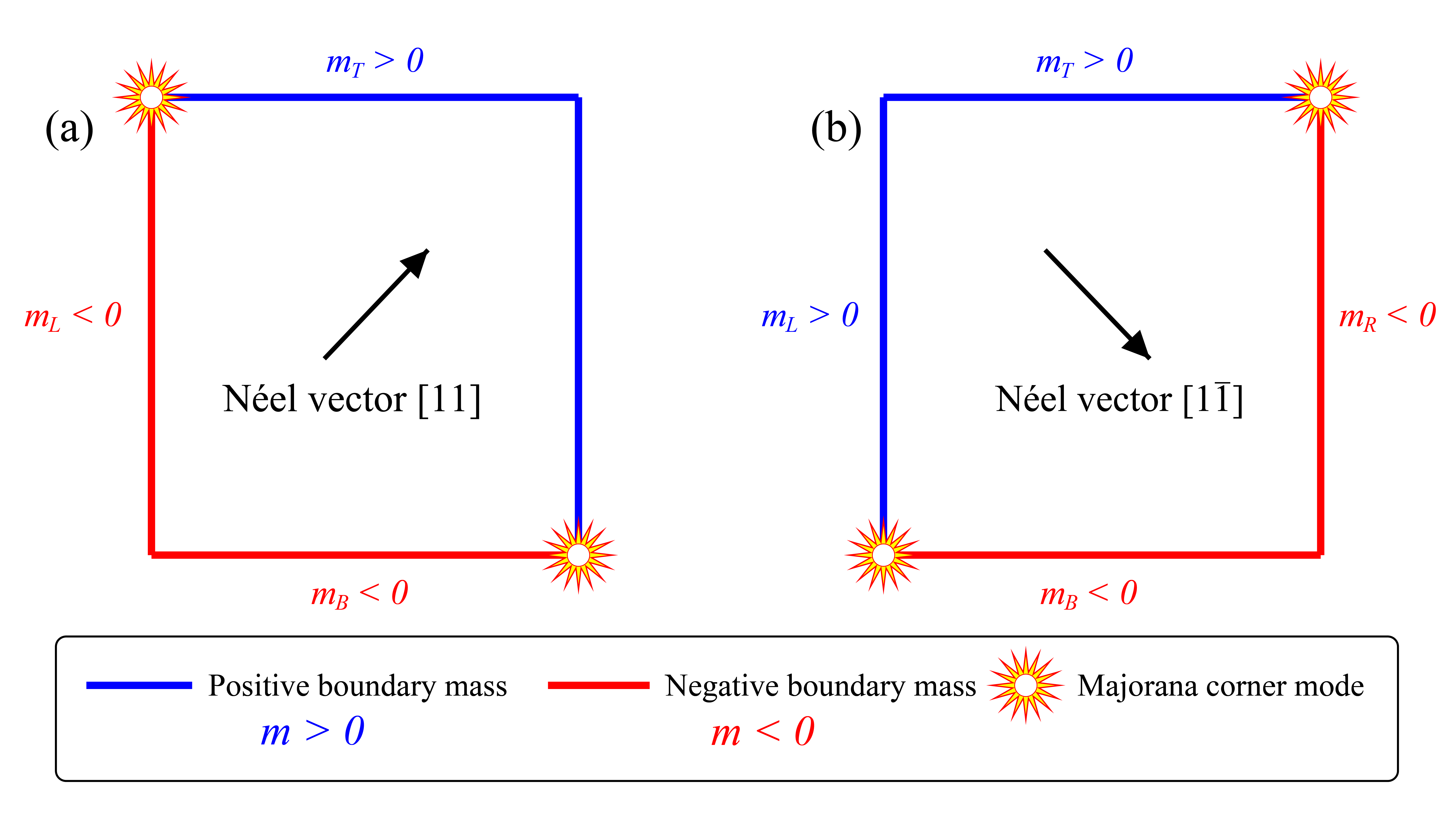} 
	\caption{MCM configurations for two N\'eel vector orientations. For a square geometry with edges at $\vartheta = 0, \pm\pi/2, \pi$ (assuming $\beta=0$), aligning $\varphi = \pi/4$ yields domain walls and MCMs at the top-left and bottom-right corners. Changing the N\'eel vector azimuth  to $\varphi = -\pi/4$ places the zero modes at the corners on the other diagonal.}
	\label{fig:corner_modes}
\end{appendixfigure}

Consequently, the projected altermagnetic perturbation introduces a mass term $H^{\text{edge}}_{\text{AM}} = m_\vartheta \chi_y$, yielding the gapped edge dispersion $E_\vartheta(k_\parallel) = \pm\sqrt{v^2 k_\parallel^2 + m_\vartheta^2}$. The anisotropic boundary mass is given by
\begin{equation}
m_\vartheta = m_0 \sin(\varphi-\vartheta) \left[ \cos\beta\cos(2\vartheta) + \frac{1}{2}\sin\beta\sin(2\vartheta) \right],
\label{eq:boundary_mass_end}
\end{equation}
where $m_0$ absorbs the  expectation value $\langle k_\perp^2 \rangle$.

Equation~(\ref{eq:boundary_mass_end}) describes how the boundary mass depends on the control parameters within the edge model. As illustrated in the polar plots of Fig.~\ref{fig:mass_polar}, $m_\vartheta$ changes sign based on the azimuthal angle $\varphi$. According to the Jackiw-Rebbi index theorem, for a finite 2D sample, the edge model predicts zero-energy modes at corners where the mass domain walls ($m_{\vartheta_1} m_{\vartheta_2} < 0$) are formed.

Consider a square geometry (boundaries at $\vartheta = 0, \pi/2, \pi, -\pi/2$) formed by a $d_{x^2-y^2}$ altermagnet ($\beta=0$). Setting the N\'eel vector to $\varphi = \pi/4$ yields the mass sequence $(m_R, m_T, m_L, m_B) = (+, +, -, -)$, trapping MCMs at the top-left and bottom-right corners (Fig.~\ref{fig:corner_modes}). Changing the N\'eel vector to $\varphi = -\pi/4$ alters the tangent spin projection, flipping specific edge masses to $(-, +, +, -)$. These parameter choices select different stationary corner configurations. Along a direct uniform rotation between them at fixed $\phi=\pi$, Eq.~(\ref{eq:boundary_mass_end}) gives vanishing mass on both ideal straight edges with $\vartheta=0,\pi$ at the intermediate orientation $\varphi=0$. This path therefore does not maintain a finite boundary mass throughout and does not establish a gapped adiabatic exchange.

\clearpage
\nocite{fisherRelationConductivityTransmission1981,lu_phi_2024,grothKwantSoftwarePackage2014,beenakkerUniversalLimitCriticalcurrent1991,iidaStatisticalScatteringTheory1990}

%

\end{document}